\documentclass{PoS}

\usepackage[utf8]{inputenc} 
\usepackage{lineno}
\usepackage{graphicx}  
\usepackage{amsmath,amssymb}  
\usepackage{subfig}
\usepackage[all=normal, tracking=tight]{savetrees}
\usepackage{url}

\usepackage[percent]{overpic}

\usepackage{natbib}
\newcommand{\aap}{Astronomy \& Astrophysics} 
\newcommand{\aapr}{Astronomy \& Astrophysics Reviews}
\newcommand{\apj}{The Astrophysical Journal}
\newcommand{\gama}{$\gamma$}
\newcommand{\goodwidth}{7.45cm}
\newcommand{\flux}{$TeV^{-1}cm^{-2}s^{-1}$}

\title{A First Look at Periodicity in HAWC with TeV Binaries}

\ShortTitle{HAWC TeV Binaries}

\author{Chad Brisbois\thanks{Corresponding Author}, Petra Huentemeyer, Henrike Fleischhack, \speaker{Binita Hona}\\
	Michigan Technological University\\ 
	Email: \email{cbrisboi@mtu.edu}, \email{petra@mtu.edu}, \email{hfleisch@mtu.edu}, \email{bhona@mtu.edu}}

\author{Chang Rho\\
        University of Rochester\\
        E-mail: \email{crho2@ur.rochester.edu}}

\author{for the HAWC collaboration \thanks{See  {\color{blue} \protect \url{http://www.hawc-observatory.org/collaboration/icrc2017.php} }for full author list.}}

\abstract{Only five binary systems have been found to emit at TeV energies. Each of these systems is composed of a massive O or B type star and a compact object (black hole or a pulsar). The type of compact object and the origin of the gamma-ray emission is unknown for most of these systems. Extending spectral observations to higher energies can help disentangle the nature of the compact object as well as the particle acceleration mechanisms present. Interestingly, the TeV emission from these systems does not always coincide with their emission in GeV or X-ray, which is how many such systems have been originally discovered. Increased coverage of these systems may allow HAWC to see precisely when in the orbit the TeV emission begins and ends. The HAWC Observatory detects TeV gamma-rays with high sensitivity, covering over two-thirds of the overhead sky every day. Applying a stacking method to known TeV binary systems can help HAWC enhance the signal from TeV binaries above the steady background from other sources in the galaxy. We will present results from this stacking analysis using 760 days of HAWC data.}

\FullConference{35th International Cosmic Ray Conference --- ICRC2017\\
        10--20 July, 2017\\
        Bexco, Busan, Korea}

\begin{document}

\section{Introduction}

Binary systems which emit \gama-rays at TeV energies are exceedingly rare objects. Only a handful are known, and are listed in Table \ref{binaryTable}. It is presently unknown whether most of these objects are driven by an accreting black hole (also called the microquasar scenario) or a pulsar wind nebula. PSR B1259-63 represents a notable exception to this trend, because it is known that the compact companion is a pulsar. The search for more of these systems is a primary focus of discovery, because both of these cases are likely to be found. It is possible that many massive binary systems go through a \gama-ray emitting phase, but due to their relatively short life as a TeV binary ($\sim10^5$ years) there are only a few in our galaxy that are observable at any given time \cite{Dubus2013}.

\begin{table}[h!]

\begin{center}
 \begin{tabular}{||c c c||} 

 \hline
 Name & Period (days) & Distance (kpc)\\ [0.5ex] 
 \hline\hline
 LS 5039 \cite{HESSBinaryReview} & 3.9  & 2.5 \\ 
 \hline
 HESS J0632+057 \cite{veritasBinary} & 315 &1.6 \\
 \hline
 LS I +61 303 \cite{LS_61_303}& 26.5* & 1.11 \\
 \hline
 HESS J1018-589 A \cite{HESSBinaryA} & 16.58  & 5 \\
 \hline
 PSR B1259-63 \cite{HESSBinaryReview}   &  1241 &  2.3 \\ 
 \hline

\end{tabular}
\end{center}
\caption{\gama-ray Binaries listed in TeVCat \href{http://tevcat.uchicago.edu/}[http://tevcat.uchicago.edu/]. Period and distance data are quoted from the references specified adjacent to their name. * LS I +61 303 has two periodicities, a further discussion of which can be found in  \cite{LS_61_303}. Of these sources, only LS 5039 is within HAWCs FoV.}  \label{binaryTable}
\end{table}

The High Altitude Water Cherenkov observatory (HAWC) is located on the Sierra Negra Volcano in the Pico de Orizaba National Park in Mexico. Due to its latitude (+19$^\circ$N), elevation (4100m), and detection method (water cherenkov detection) it is sensitive to TeV emission on an extremely high duty cycle across 2/3 of the sky each day \cite{crabPaper}. This high uptime ($>$90$\%$) allows HAWC to be sensitive to time variability in emission that other instruments cannot observe. HAWC has a real-time flare monitoring system which is capable of detecting flares on time scales as small as two minutes \cite{flarePaper}. These qualities make HAWC an excellent instrument for observing time-dependent behavior in the TeV sky. Presently HAWC uses what is referred to as fractional nHit bin ($0 \leq \mathcal{B} \leq 9$) analysis \cite{crabPaper}, where energy proxy bins are defined by the fraction of the detector triggered by a particular air shower.  

\begin{figure}[h!]
   \centering
   \begin{overpic}[width=12cm,tics=10]{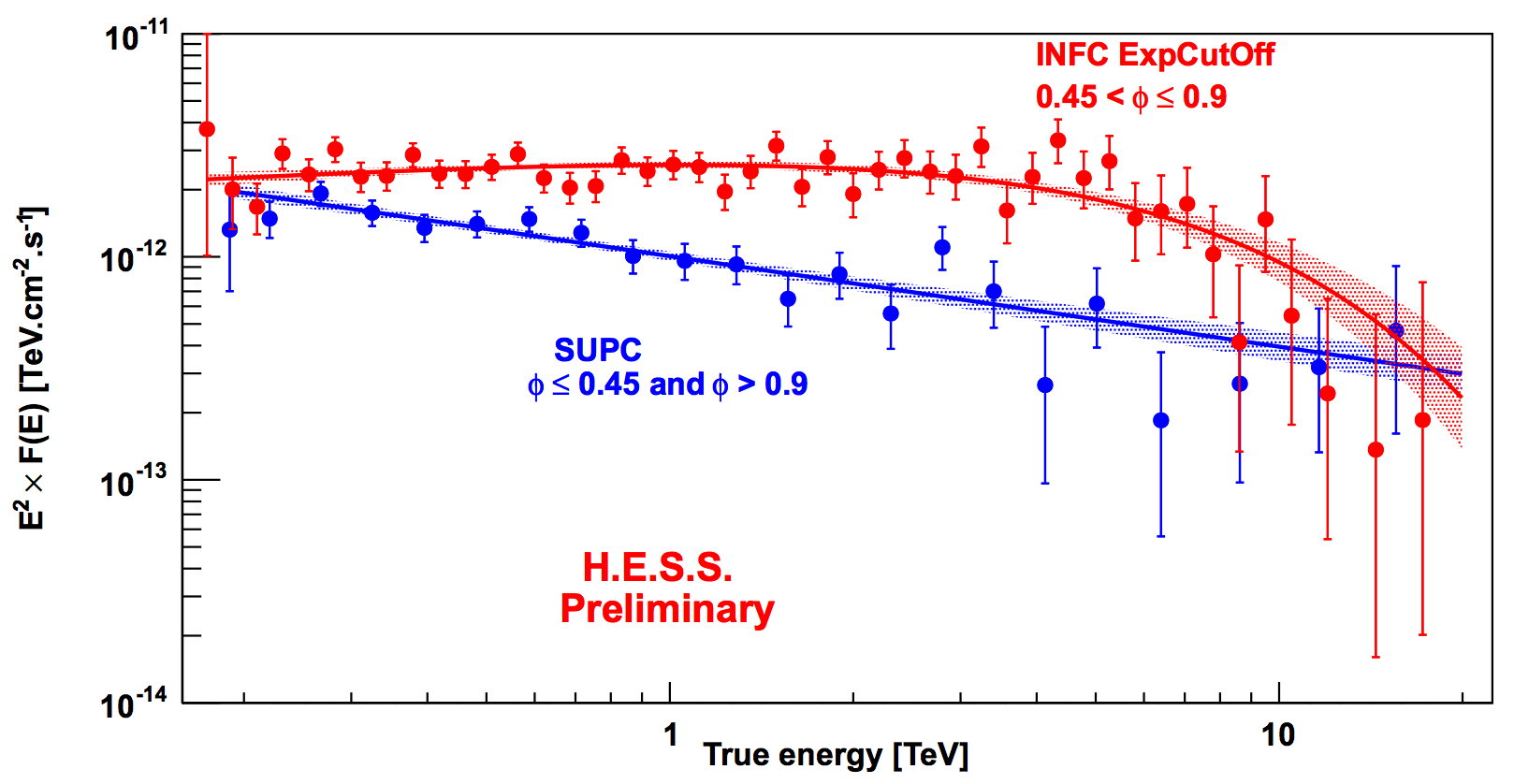}\end{overpic}
   \caption{Spectrum of LS 5039 binned into high and low state, taken from \cite{HESSBinaryReview}.}
   \label{fig:sedMariaud}
\end{figure}

LS 5039 is a high-mass X-ray binary system composed of a massive O-type star ($\textrm{M}=23\textrm{M}_\odot$) and a compact object ($\textrm{M}=3.7\textrm{M}_\odot$) with an orbital period of 3.9 days \cite{HESSBinaryReview}. The nature of the compact companion is unknown. A variety of instruments such as H.E.S.S. and Fermi-LAT have detected \gama-ray emission from LS 5039. In \cite{ls5039paper}, H.E.S.S. was able to observe two distinct spectral shapes in the \gama-ray emission of LS 5039 depending on the orbital phase ($0 < \phi \leq 1$) of the system. In the high state ($0.45 <\phi\leq 0.9$), LS 5039 exhibited a hard power law spectrum with a cutoff, and in the low state ($0.45 \leq\phi\, \&\,\phi > 0.9$) a simple power law with a softer index. A plot of the spectral energy distribution can be found in Figure \ref{fig:sedMariaud} which was taken from \cite{ls5039paper}. The same definitions of high and low state are used in the present analysis.

From Figure \ref{fig:sedMariaud} it is ambiguous what the spectrum beyond 10 TeV looks like. If the high state truly cuts off beyond this energy range, or if the high state emission is simply "on top of" the low state emission is an open question that HAWC may be able to address. HAWC is currently the only instrument that has hundreds of days of exposure time on sources at energies greater than 10 TeV.

\begin{figure}[h!]
   \centering
   \begin{overpic}[width=10cm,tics=10, trim={0 35pt 0 47pt}, clip]{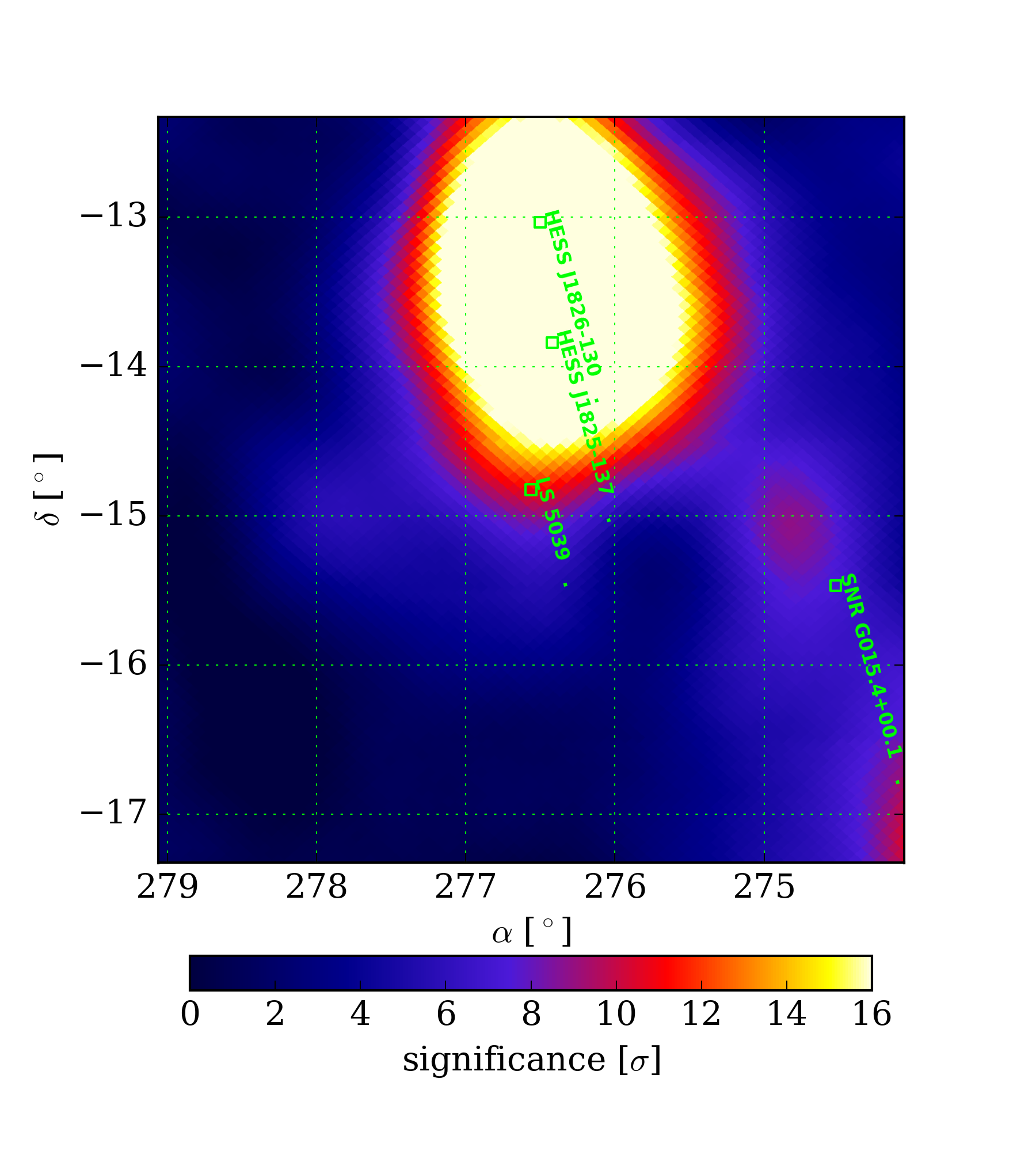}
   \put(20,30){\color{white} HAWC Preliminary }
   \end{overpic}
   \caption{Significance map of the region around LS 5039 using 760 days of accumulated HAWC data.}
   \label{fig:760daymap}
\end{figure}

LS 5039 is embedded in a busy region along the galactic plane. There are two bright extended sources nearby, HESS J1825-137 and HESS J1826-130. HAWC detects a significant \gama-ray excess over the expected background at the location of LS 5039 (see Figure \ref{fig:760daymap}). However, it is not clear how much of this excess is due to \gama-ray emission from LS 5039 and how much is due to leakage from the other two sources. In this presentation, we will explore a method to measure the difference in flux between the high and low state of LS 5039 as seen by HAWC.

\section{Method}
HAWC events are binned according to the phase of LS 5039. The data is analyzed with the standard HAWC analysis software  \cite{crabPaper} \cite{2hwcCatalog}. The two flux maps are produced by fitting a single point source with a power-law spectrum at the position of each pixel, with the spectral index fixed to -2.0. The quoted fluxes refer to the fitted differential flux at the pivot energy of 7 TeV.  The high state map contains 342 days of data, and the low state contains 418 days.



By further subdividing the two sets of events for the high and low states of the binary into bins corresponding to the low-energy nHit bins and the high-energy nHit bins, we can gain a small window into the relative flux between the two states. If there is a positive relative flux in the low energy bins, and a negative relative flux in the high energy bins, then that is consistent with the high state flux not simply being "on top of" the low state flux. If at high energies the difference is zero, then it possible that the process generating the high state turns off at those energies, but the low state process persists.




 Assuming the uncertainty on the flux normalization is Gaussian and purely statistical, the significance of the flux difference between the high and the low state maps is estimated by Equation \ref{eq:relFlux}. 
 The difference between the flux in the high ($F_{high}$) and low($F_{low}$) states gives an estimate of the significance by dividing by the uncertainty ($\sigma$)in the measured flux at high and low energies.
 

\begin{align} 
 \label{eq:relFlux}
    \mathfrak{S} = {}& \frac{F_{high} - F_{low}}{\sqrt{\sigma_{high}^2 + \sigma_{low}^2}} 
 \end{align}

\begin{figure}[htp!]%
     \subfloat[Crab Nebula Map]{{ 
     \begin{overpic}[width=\goodwidth, trim={0 1.2cm 0 1.70cm}, clip, tics=10]{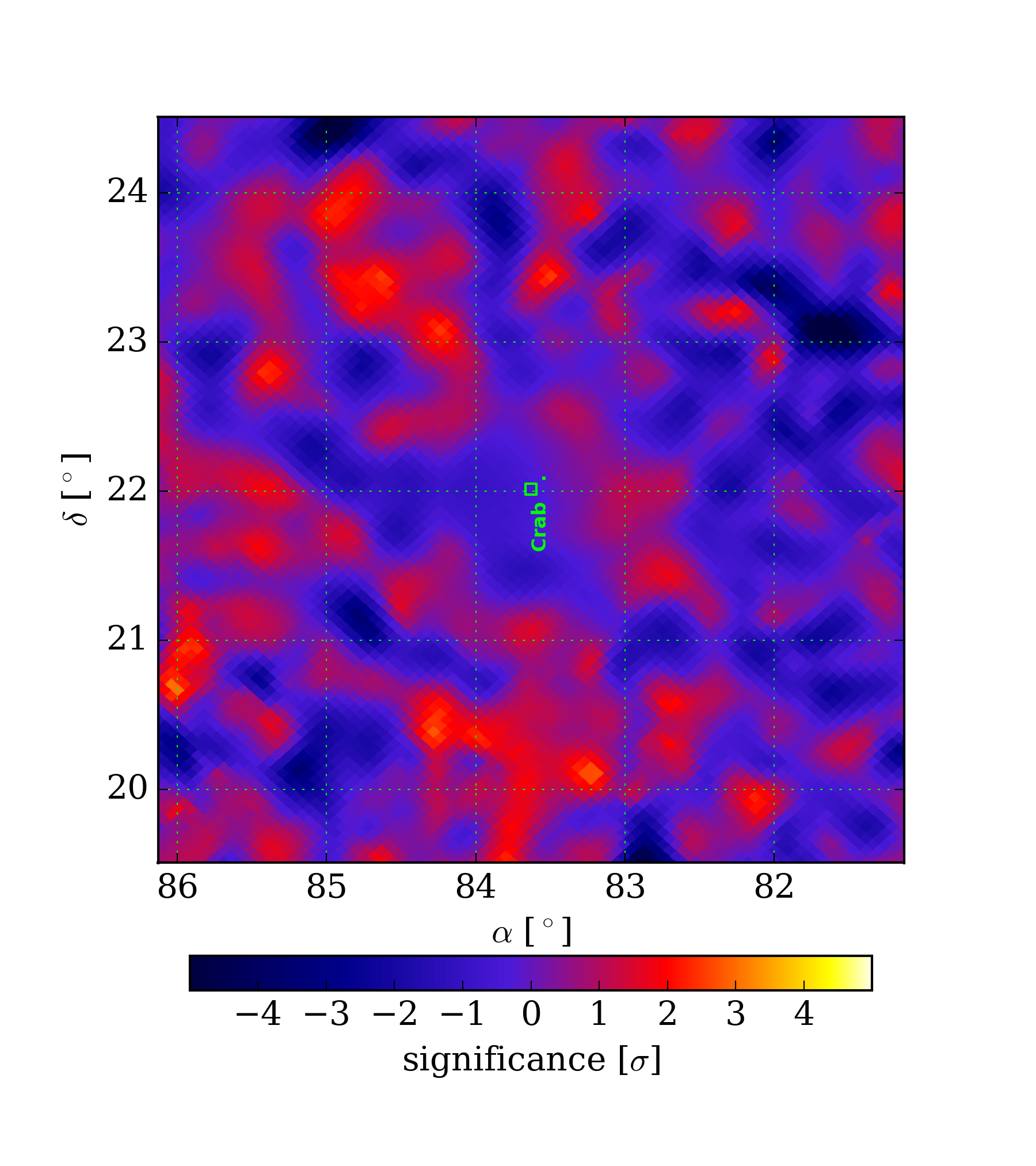}
   \put(20,30){\color{white} HAWC Preliminary }
   \end{overpic}
     }}
     \subfloat[Crab Nebula Distribution]{{
   \begin{overpic}[width=\goodwidth, trim={0 -70pt 0 33pt}, clip, tics=10]{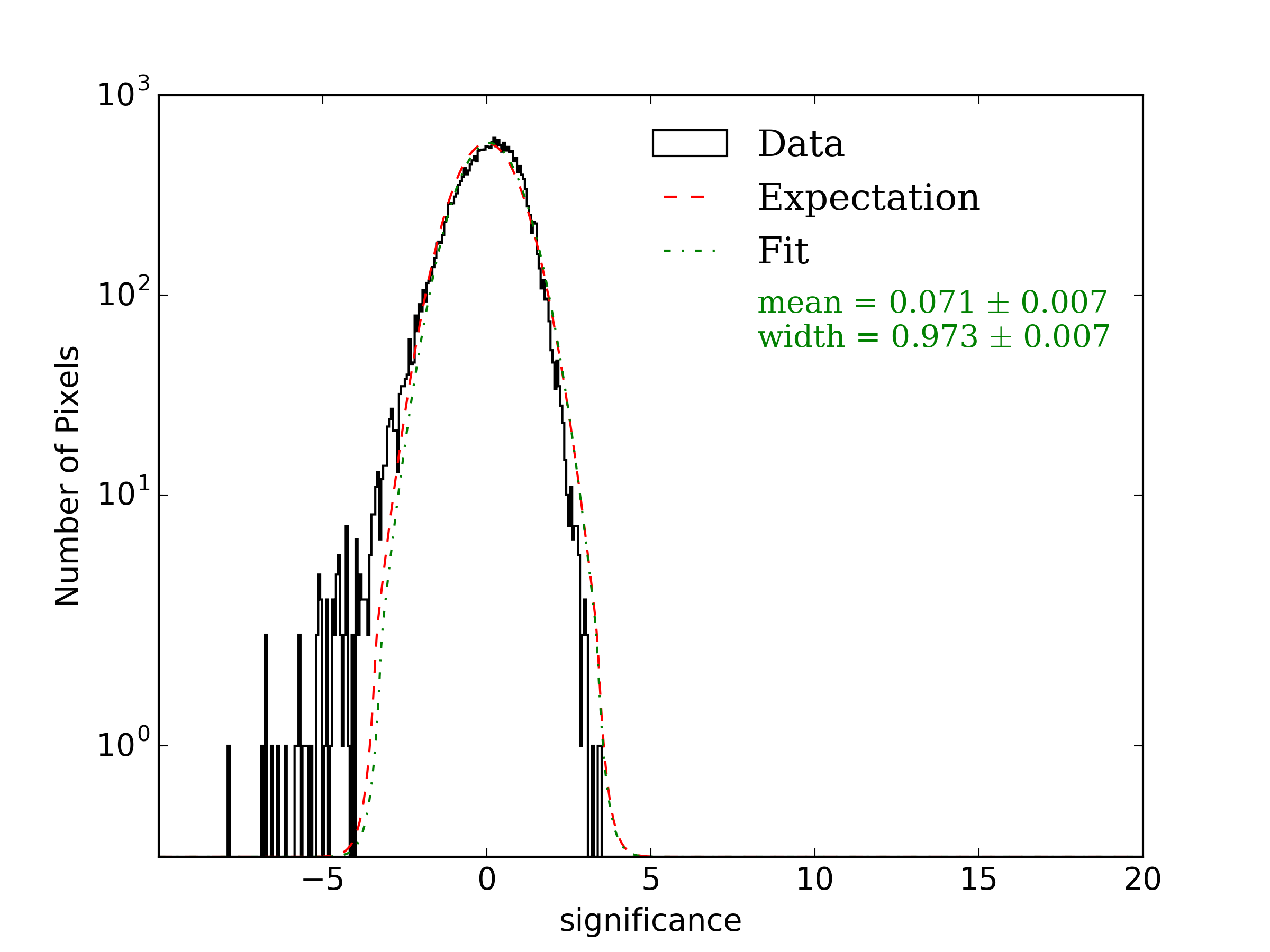}
   \put(55,35){\color{black}HAWC}
   \put(52,30){\color{black}Preliminary}
   \end{overpic}
     }}
    \caption{Significance of flux differences in Crab Nebula Region.}%
    \label{fig:crabGone}%
\end{figure}

 This procedure is tested by applying it to a non-periodic source, the Crab nebula. Applying Equation \ref{eq:relFlux} to the high and low state maps, the Crab nebula has disappeared from the maps, see Figure \ref{fig:crabGone}. The result of this important check is to be expected \cite{crabPaper}, considering that the periodicity is fixed to the periodicity of LS 5039. In Figure  \ref{fig:crabGone}b, we see that the distribution is indeed approximately Gaussian, and therefore $\mathfrak{S}$ is treated as significance.
 
\section{Results}

\begin{figure}[h!]
   \centering
   \begin{overpic}[width=8cm, trim={0 35pt 0 47pt}, clip, tics=10]{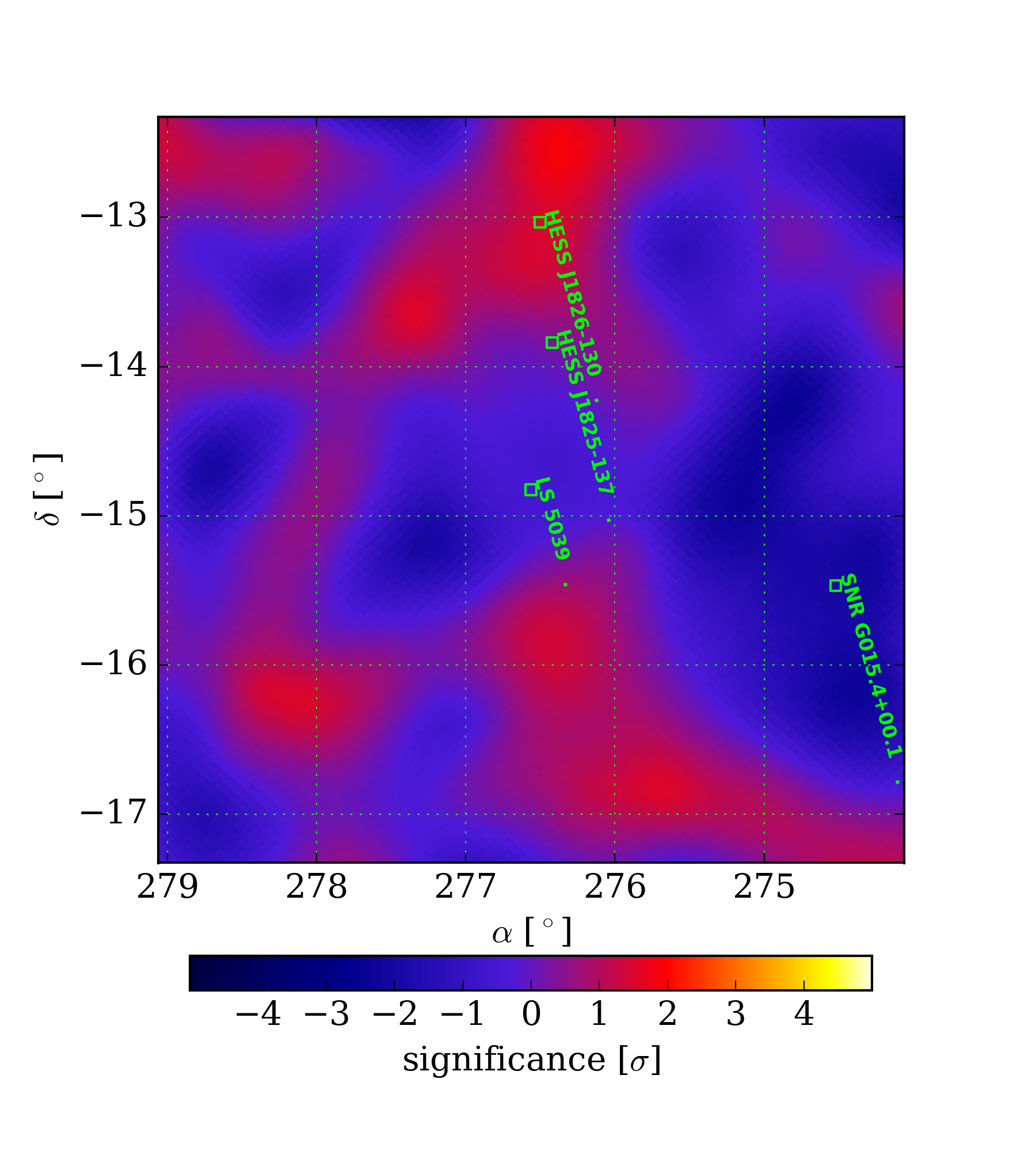}
   \put(20,30){\color{white} HAWC Preliminary }
   \end{overpic}
   \caption{Overall relative flux between high and low state maps.}
   \label{fig:allBinsFluxDiff}
\end{figure}

In the region around LS 5039, the high and the low state maps display structure that is similar to what is seen in the full data set (see Figure \ref{fig:760daymap}), with nearby HESS sources dominating the emission of the region. The morphology of HESS J1825-137 and HESS J1826-130 are not known precisely enough to disentangle their emission from any eventual signal from LS 5039.  Therefore we attempt to distinguish the relative flux between the two states, doing a per pixel subtraction of the flux according to equation \ref{eq:relFlux}. Following this prescription, the high and low state maps, can be converted to the relative flux between the states, as seen in Figure \ref{fig:allBinsFluxDiff}. The fluxes at the location of LS 5039 are listed in Table \ref{resultsTable}. No significant detection of a difference between the high and low state of LS 5039 is observed, nor are there significant detections elsewhere in the region.


\begin{table}[htp!]
\begin{center}
 \begin{tabular}{||c c||} 
 \hline
 Map & Flux ($10^{-14}$ \flux)  \\ [0.5ex] 
 \hline\hline
 High State  & $1.13\pm 0.25$  \\ 
 \hline
 Low State  & $1.32\pm 0.24$  \\
 \hline
 Difference &  $-0.19 \pm 0.35$\\
 \hline
\end{tabular}
\end{center}
\caption{Differential flux at 7 TeV measured at the location of LS 5039 in the 'high' and 'low' states as well as the flux difference between the two states. Spectral fit was performed over the full energy range. Statistical uncertainties only. The quoted flux values have not been corrected for leakage from HESS J1825-137 and HESS J1826-130 and should not be taken as a measurement of the flux from LS 5039.}  \label{resultsTable}
\end{table}

\begin{figure}[h!]%
    \centering
    \subfloat[Low energy sample ($2\leq \mathcal{B} \leq 5$) ]{{
     \begin{overpic}[width=\goodwidth, tics=10, trim={0 54pt 0 40pt}, clip]{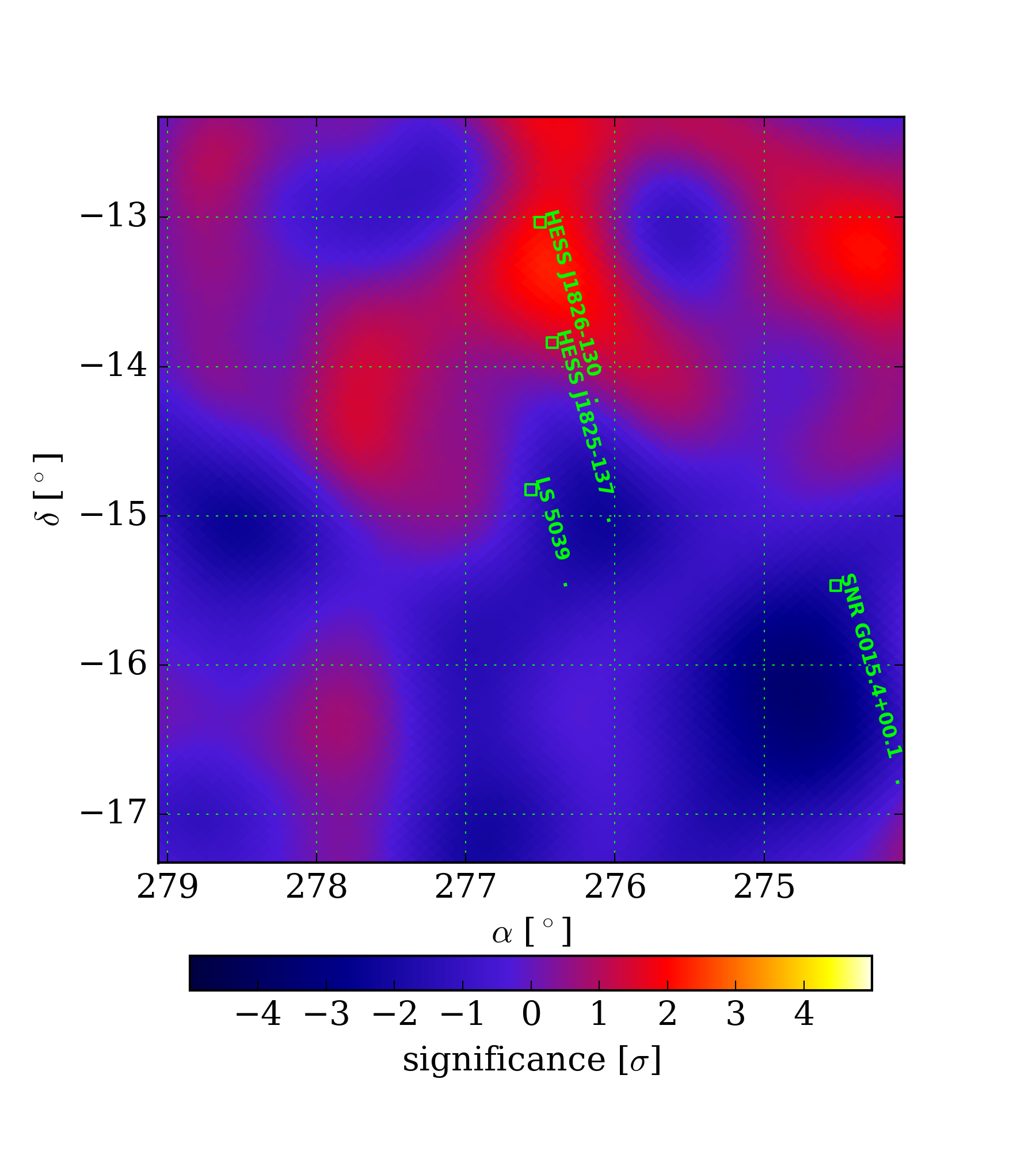}
   \put(20,30){\color{white} HAWC Preliminary }
   \end{overpic}
    }}%
    \subfloat[High energy sample  ($6 \leq \mathcal{B} \leq 9$) ]{{
     \begin{overpic}[width=\goodwidth, tics=10, trim={0 54pt 0 40pt}, clip]{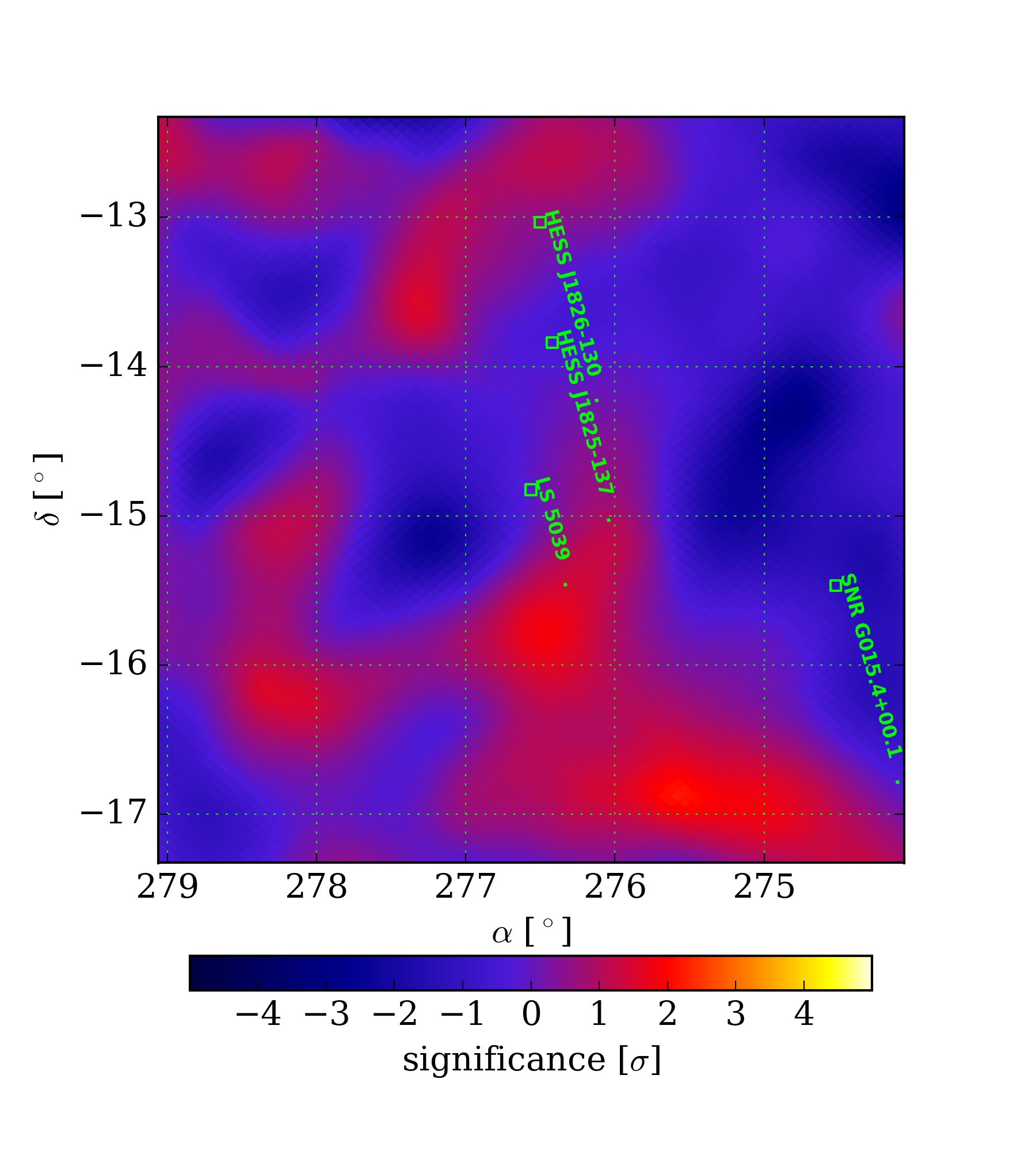}
   \put(20,30){\color{white} HAWC Preliminary }
   \end{overpic}
    }}%
    \caption{Significance of flux differences in high and low state maps in the region around LS 5039}%
    \label{fig:compare}%
\end{figure}

Splitting into the high and low energy maps for the relative flux between the states, corresponding to the crossover point near 10 TeV, there is again no significant difference between the two flux states (See Figure \ref{fig:compare}). All variations are consistent with statistical fluctuations. The flux measurements at LS 5039 are summarized in Table \ref{resultsTable2}. One critical caveat to the flux measurements listed in Tables \ref{resultsTable} \& \ref{resultsTable2} is that they contain any emission from LS 5039 as well as contamination from other sources.

\begin{table}[htp!]
\begin{center}
 \begin{tabular}{||c c c||} 
 \hline
 Map                  & Flux Difference ($10^{-15}$ \flux)   & $\mathfrak{S}$ \\ [0.5ex] 
 \hline\hline
 High nHit bins ($6 \leq \mathcal{B} \leq 9$)  & $-0.59 \pm 3.2$  &  $-0.18$ \\ 
 \hline
 Low nHit bins ($2\leq \mathcal{B} \leq 5$)   & $-6.2 \pm 8.4$    & $-0.75$ \\
 \hline
 All nHit bins ($2\leq \mathcal{B} \leq 9$)      &  $-2.0 \pm 3.5$ & $-0.54$ \\
 \hline
\end{tabular}
\end{center}
\caption{Flux differences at the location of LS 5039, along with significance ($\mathfrak{S}$) for the full energy range as well as $2< \mathcal{B} < 5$ and $6 < \mathcal{B} < 9$.}.  \label{resultsTable2}
\end{table}

\section{Discussion}
We have presented a method to measure the flux differences between the high and low states of LS 5039. The method was validated by applying it to the crab nebula, a strong \gama-ray sources that does not exhibit periodic variability. In Figures \ref{fig:allBinsFluxDiff} \& \ref{fig:compare}, the relative flux was consistent with zero in every case. This analysis does not detect any significant difference in the flux between the high and low states for LS 5039. However, the systematics of the method itself are being investigated, with the intention of producing upper limits in the future.  As HAWC transitions from a fractional nHit bin based analysis to an analysis using energy estimators for measuring the spectrum, the ability to make precise statements about the flux (and variation thereof) at the location of LS 5039 or other variable sources for different energy bins will make HAWC more sensitive to differences in the differential flux.
Additionally, efforts to measure the \gama-ray emission from LS 5039 by modeling the morphology of the two neighboring sources are currently underway. Planned improvements in HAWC's angular resolution due to improvements in shower reconstruction as well as the planned outrigger array will also help disentangle the emission in this complex region.  



\section{Acknowledgements}
We	acknowledge	the	support	from:	the	US	National	Science	Foundation	(NSF);	the	
US	Department	of	Energy	Office	of	High-Energy	Physics;	the	Laboratory	Directed	
Research	and	Development	(LDRD)	program	of	Los	Alamos	National	Laboratory;	
Consejo	Nacional	de	Ciencia	y	Tecnolog\'{\i}a	(CONACyT),	M{\'e}xico	(grants	
271051,	232656,	260378,	179588,	239762,	254964,	271737,	258865,	243290,	
132197),	Laboratorio	Nacional	HAWC	de	rayos	gamma;	L'OREAL	Fellowship	for	
Women	in	Science	2014;	Red	HAWC,	M{\'e}xico;	DGAPA-UNAM	(grants	RG100414,	
IN111315,	IN111716-3,	IA102715,	109916,	IA102917);	VIEP-BUAP;	PIFI	2012,	
2013,	PROFOCIE	2014,	2015; the	University	of	Wisconsin	Alumni	Research	
Foundation;	the	Institute	of	Geophysics,	Planetary	Physics,	and	Signatures	at	Los	
Alamos	National	Laboratory;	Polish	Science	Centre	grant	DEC-2014/13/B/ST9/945;	
Coordinaci{\'o}n	de	la	Investigaci{\'o}n	Cient\'{\i}fica	de	la	Universidad	
Michoacana. Thanks	to	Luciano	D\'{\i}az	and	Eduardo	Murrieta	for	technical	
support.
C. Brisbois acknowledges support under the U.S. Department of Energy, Office of Science, Office of Workforce Development for Teachers and Scientists, Office of Science Graduate Student Research (SCGSR) program.

\end{document}